\documentclass[12pt,preprint]{aastex}
\usepackage{longtable}

\shorttitle{The Properties of Satellites in External Systems}
\shortauthors{C. M. Gutierrez \& M. Azzaro}

\begin{document}

\title{The Properties of Satellite Galaxies in External Systems. 
II. Photometry and Colors}
\author{Carlos M. Guti\'errez} \affil{Instituto de Astrof\'{\i}sica de Canarias,
E-38205 La Laguna, Tenerife, Spain} \email{cgc@ll.iac.es} 
\and \author{Marco Azzaro} \affil{Isaac Newton Group of Telescopes, Ap.
321, E-38700 S/C de La Palma, La Palma, Spain}

\begin{abstract} 

In this  second paper dedicated to the study of satellite galaxies we present
broad-band photometry in the $B$, $V$, $R$ and $I$ filters of 49 satellite
galaxies orbiting giant isolated spiral galaxies. First analysis of the
properties of these objects are presented by means of color--color and
color--magnitude diagrams for early- and late-type satellites.  Although we
find differences in the slope of the $V-I$ vs. $M_v$ color magnitude diagram, as
a whole, the relations are in agreement with the trends known to date for
galaxies of similar magnitudes in nearby clusters of galaxies. Comparison with
the relations found for satellites in the Local Group allows us to sample
better the bright end of the luminosity function of satellite galaxies and
extends for brighter objects the validity of the color--magnitude relation
found for dwarf galaxies  in the Local Group. Most of the E/S0 galaxies in our
sample show a negative color gradient with values similar to those known for
early type galaxies in other environments.

\end{abstract}


\section{Introduction}

The study of the small scale structure in the Universe is a much discussed
topic in current astronomy. One of the reasons for this wide interest is 
the discrepancies  between observations and the predictions of the standard
cold dark matter (CDM) model. For instance, from $n$-body simulation (Moore et
al.\ 1999; Klypin et al.\ 1999) the expected  number of satellites orbiting
galaxies like the Milky Way or Andromeda is an order of magnitude larger than
that observed in the Local Group (e.g., Mateo 1998). This is a potentially 
strong objection against the hierarchical scenario proposed in the standard
CDM model. To solve this discrepancy several mechanisms that suppress the
formation of satellites after the re-ionization  in the early epoch of the
Universe have been proposed (e.g., Bullock, Kravtsov, \& Weinberg 2000). So
far, it  is unclear whether this or any of the other mechanisms proposed is
able to reconcile fully the predictions from the models and the observations.

Satellite galaxies are also interesting for tracing the gravitational
potential and   for estimating the mass of the parent galaxy at
distances unreachable with other methods (Erickson, Gottesman \& Hunter
1999). The standard models predict a decline in satellite galaxy
velocity with distance to the primary. This has been observationally
explored by Zaritsky et al.\ (1997) and Prada et al.\ (2003). The
predicted effect was not found in the analysis of the first  group.
However, Prada et al., with better statistics and the removal of
interlopers,  have claimed the existence of this decline in the way
predicted by the standard models.

Our knowledge of satellite galaxies beyond the Local Group is still very limited, owing
 to the intrinsic faintness of these objects. For instance, in the
Local Group the known dwarf satellite galaxies have brightnesses in the range $-18\le M_B
\le -8$. So detailed studies of external systems are limited
to galaxies in the nearby Universe. Even surveys such as the Sloan Digital Sky
Survey (Prada et al.\ 2003) are  capable to sampling only the very bright part of the
luminosity function of these objects, typically detecting   1--2
satellites orbiting each giant galaxy. For instance, the external isolated
galaxies in which a larger number of galaxies have been found are NGC~1961 (Gottesman,
Hunter \& Shostak 1993) and NGC~5084 (Carignan et al.\ 1997) in which 5 and 9
satellites respectively have been catalogued.

Bearing all the above points in mind, we decided to start a study of satellite
galaxies orbiting external isolated galaxies. We extracted our  sample from
the compilation of  such objects by Zaritsky et al.\ (1997). Only spiral
parents were included in that catalogue. A similar catalogue for parents with
elliptical morphologies is being compiled by Smith \& Martinez (2003). With
photometry in optical broad ($BVRI$) and narrow bands  (H$\alpha$), and near
infrared ($J$ and $K$) of both the satellites and their parent galaxies we
address the following questions: i) what are the properties (statistics,
luminosity function,  morphology, structure, colors, etc.) of the satellite
galaxies, and how do they compare with those found in the Local Group;  ii)
how  the above properties are related to the relative positions of the
satellite galaxies with respect to their parents; iii) what  the possible
interactions are between satellites, and between satellites and their 
parents; and iv) the star formation rate and how it is related with the
overall properties and relative position of the satellites. In the first paper
of this series, Gutierrez, Azzaro, \& Prada (2002, hereafter Paper~I), we
presented the structural parameters and morphological classification of $\sim
60$ of such objects.  Here, we present the photometry and first analysis of
the colors of most of them. A full analysis combining structure and photometry
and testing against the predictions of theoretical models will be conducted
and presented in the near future.

\section{Observations and Photometry}

The observations presented here were performed during several runs at the
IAC80,\footnote{The IAC-80  is located at the Spanish Teide Observatory on the island
of Tenerife and is operated by the Instituto de Astrof\'\i sica de Canarias.}
the Nordic Optical Telescope (NOT),\footnote{The Nordic Optical Telescope  is operated on the island of La
Palma jointly by Denmark, Finland, Iceland, Norway, and Sweden, in the Spanish
Roque de los Muchachos of the Instituto de Astrof\'\i sica de
Canarias Observatory.}, and the 1.23 and 2.2 m Calar Alto\footnote{The German-Spanish Astronomical
Centre, Calar Alto, is operated by the Max-Planck-Institute for Astronomy,
Heidelberg, jointly with the Spanish National Comission for Astronomy.}
telescopes. The details of the sample can be found in Zaritsky et al.\ (1997)
and Paper~I. Typical  integration times per filter and galaxy range from 
$\sim 30$ to $\sim 90$
minutes on the IAC80, and from $\sim$5 to $\sim$30 min on the other
telescopes. The seeing was between 1.5 and 2.0 arcsec in most  cases.
The observations were usually performed in photometric conditions and
calibrated using Landolt  stars (Landolt 1992). In a few cases of 
non-photometric nights we made relative calibrations with photometric nights using
bright field stars (typically 4--8 in each case). The typical rms accuracy of
the calibration is $\sim$0.03 mag. The data were reduced using
IRAF.\footnote{IRAF is the Image Reduction and Analysis Facility, written and
supported by the IRAF programming group at the National Optical Astronomy
Observatory (NOAO) in Tucson, Arizona.}  More details on the observations and
reduction can be found in Paper~I. The magnitudes presented here were
computed using the software Sextractor  (Bertin \& Arnouts 1996) and correspond
to the total integrated magnitudes computed in apertures defined in the way proposed by Kron
(1980). The  uncertainty in the estimation of these magnitudes are dominated by
the sky subtraction and the extrapolation of the external profile of the
galaxies. We estimate the typical overall uncertainty as $\sim$0.1 mag.

Table~1 presents the integrated photometry in the four bands. We have
photometry for 49 objects, $\sim$80\% of them in all four bands.
For most of them we have also analyzed the morphology and structure (see Paper~I).
The galaxies span a range of $\sim$6 mag in apparent magnitude. When converted
to absolute magnitudes, this range is $\sim$5 mag, i.e., a factor of $\sim$100
in mass.

\begin{figure}  
\begin{center}
\includegraphics[width=12cm,angle=0]{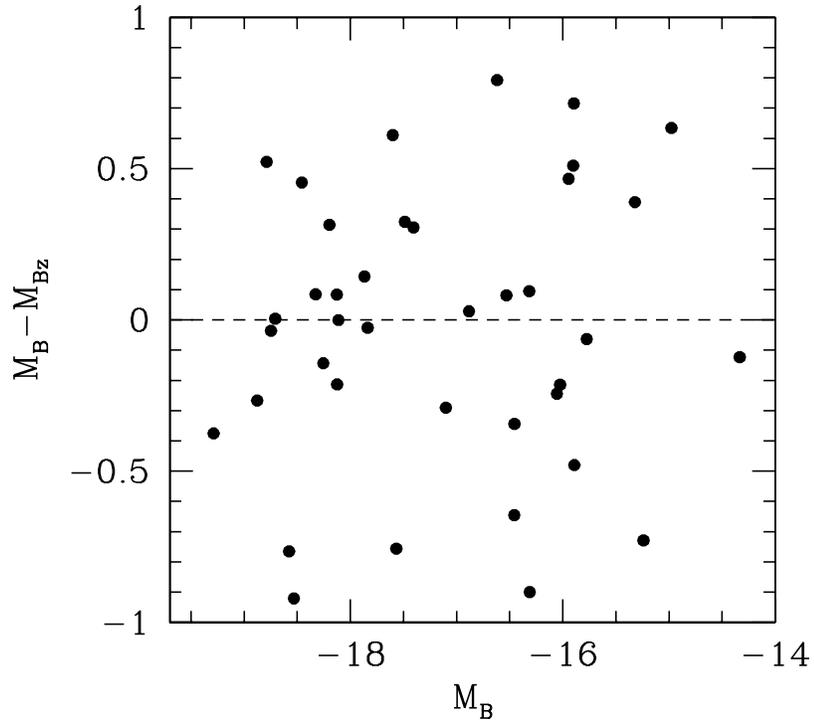}
\protect\caption[ ]{Comparison between the absolute magnitudes derived by Zaritsky et
al. (1997) ($M_{Bz}$) and those ($M_B$) derived from the observations presented in
this paper.} 
\label{comparison}
\end{center}
\end{figure}

We have made a comparison between the $M_B$ magnitudes presented by
Zaritsky et al.\ and our estimate in this band. To obtain the absolute
magnitudes from our measurements we have used a value of the Hubble
constant of 72 km s$^{-1}$ Mpc$^{-1}$. We have excluded in this
comparison the object NGC~5965$a$ because we have found two objects
(denoted as NGC~5965$a$1 and NGC~5965$a$2 in Table~1) that are very
close together (the angular distance between them is 1.6 arcsec), and it
is unclear if the magnitude quoted by Zaritsky et al.\ corresponds to
one of the two components or to the combination of both. The objects
show evidence of interaction, so we think that both are satellites of
NGC~5965. After these objects were excluded, we have 39 objects with
measurements in the $B$ band. The absolute magnitude in the $B$ band 
Zaritsky et
al.\ (denoted here by $M_{Bz}$) were slightly modified in order to take into account the different
values of the Hubble constant used by these authors and by us (75 and 72
km s$^{-1}$ Mpc$^{-1}$ respectively). In Figure~\ref{comparison} we
present this comparison. The differences are $\Delta =|M_B-M_{Bz}|\le
1$. The figure shows no systematic error,  the  mean value of the
residuals being  0.31 mag and the rms $0.45$ mag. Considering that the
magnitudes reported by Zaritsky et al.\ are photographic, and that these
authors estimated a proper uncertainty of $\pm 0.5$ mag, we conclude
that both estimates are in very good agreement. This comparison also
indicates that our uncertainties in the estimation of the magnitudes are
small compared with the above value.

\section{The Color--Magnitude Relation for Satellite Galaxies}

The existence of a color--magnitude relation for  elliptical galaxies
in clusters is widely known (e.g.,  Visvanathan \& Sandage 1977). The
origin of this relation is controversial: while some authors (Kodama
\& Arimoto 1997) have argued that the relation is a consequence of
changes in metallicity, others (Ferreras, Charlot, \& Silk 1999) think
that it is a consequence of changes in both age and metallicity.
Recently (Vazdekis et al.\ 2001) have analyzed high signal-to-noise
ratio spectra of six elliptical galaxies in the Virgo Cluster using a
new spectral index, and conclude that the color--magnitude relation is
a consequence of a relation between luminosity (or mass) and
metallicity. For the analysis presented in this section, we have
corrected the magnitudes quoted in Table~1 for Galactic extinction 
using the model by Schlegel, Finkbeiner, \& Davis (1998).

\begin{figure}  
\begin{center}
\includegraphics[width=12cm,angle=0]{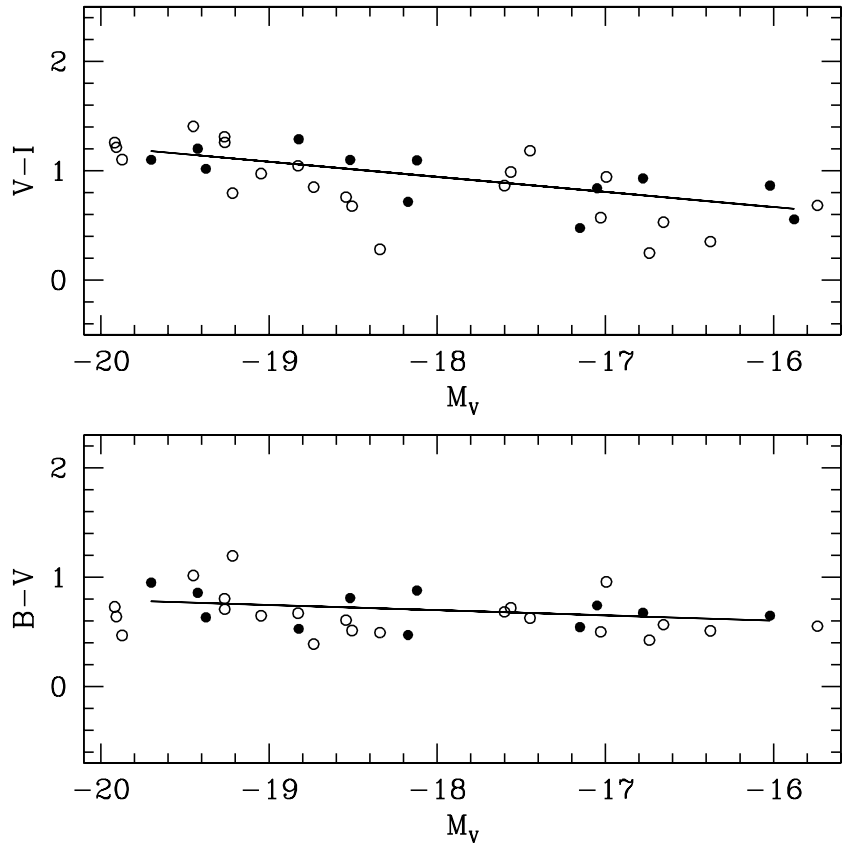}
\protect\caption[ ]{Color--magnitude diagrams for early type ($filled \;
circles$) and late type ($open \; circles$) satellite galaxies. The lines are
least-squares fits to the early type galaxies.} 
\label{cmr}
\end{center}
\end{figure}

\begin{figure}  
\begin{center}
\includegraphics[width=12cm,angle=0]{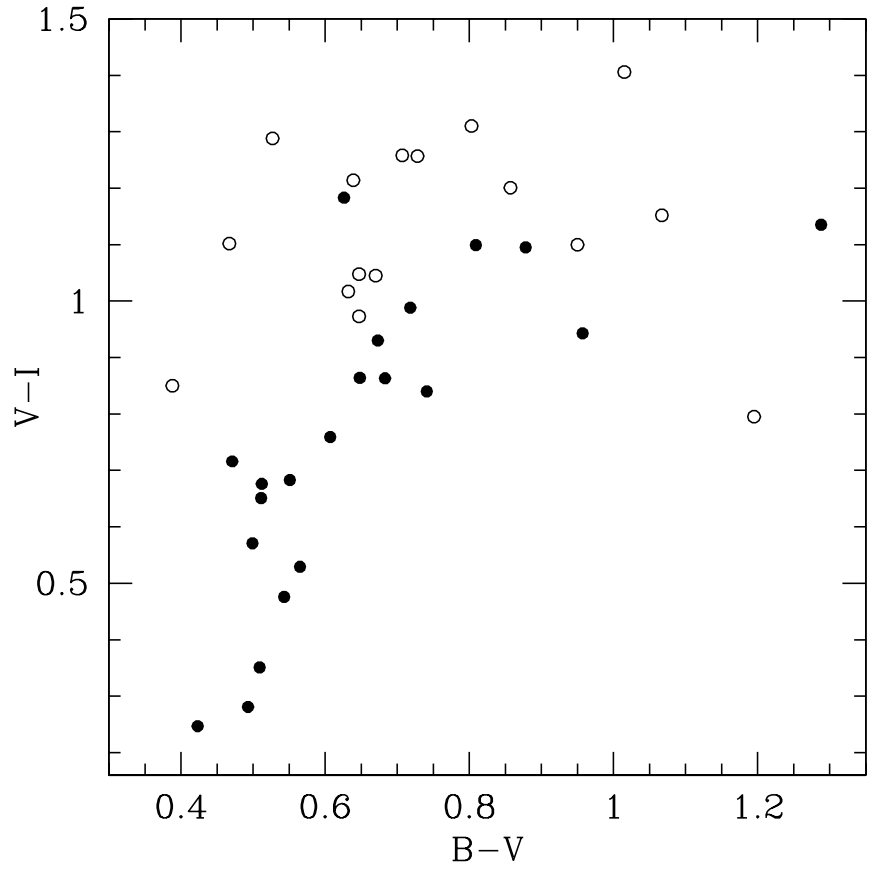}
\protect\caption[ ]{Color--color diagram of the satellite galaxies with $M_B\le
-18$ ($open\; circles$) and $M_B\ge -18$ ($filled\; circles$).} 
\label{col}
\end{center}
\end{figure}

Figure~\ref{cmr} presents a color--magnitude diagram for the galaxies
classified as early and late types in our sample. The objects span a range of
$\sim$4 mag while they are in a narrow range in the $B-V$ and $V-I$  colors. 
Although the  statistics are poor because of the low number of objects
analyzed,  the existence of a tight relation between colors and magnitudes is
clear. This color--magnitude relation is similar for both types of galaxies
(early and late), although it is notable that the  dispersion of the early
types with respect to these relations is smaller. The mean values are +0.70 and
+0.65 for the $B-V$ colors, and  $+0.93$ and $+0.88$ for the $V-I$ colors of early
and late types respectively.  For these two type of galaxies we have conducted
least squares fits of the equations  $B-V=aM_V+b$ and $V-I=aM_V+b$
respectively. For these fits we assigned the same weight to all galaxies. The parameters of these fits  are  presented in Table~2.
Combining the two colors $B-V$ and $V-I$, we obtain the color--color diagram
presented in Figure~\ref{col}. As expected from the previous figure, galaxies
tend to occupy a narrow region in this diagram, with the  brightest galaxies
being redder than the faintest ones. Roughly speaking, the  position (0.8, 1.0)
in the  ($B-V$, $V-I$) plane separates galaxies with  $M_B\le -18$ from those
with $M_B\ge -18$.  We have compared these results with those found for
galaxies with similar magnitudes in the Fornax Cluster by Karick, Drinkwater, 
\& Gregg  (2003), and Griersmith (1982). We  note similar trends but with a
larger dispersion in our objects; and  a reddening  with luminosity similar in
the $B-V$ colors and larger in $V-I$ for the galaxies of our sample. For
instance, Karick et al. obtained mean values of $B-V$ of 0.70 and 0.57 for
their sample of dwarf  ellipticals and late types. These values are roughly
compatible with those that we have obtained. The small discrepancy found
between the $B-V$ colors of late types  is partially due to the different
magntitude range of galaxies in both samples. Karick et al. obtained a 
slope for the $B-V$ relation of the early type population of $-0.034\pm 0.006$;
this and the value found by Griersmith (1982) are compatible with our results. 
However the slope obtained for the $V-I$ curve is different between the Fornax
members and the objects analyzed here. The slope of this color-magnitude relation
is  larger in our sample. If we restrict our analysis to the
common range in magnitude between both samples, this difference persists.  
Possible reasons for this  difference  will be analyzed in a future paper. 

\begin{figure}  
\begin{center}
\includegraphics[width=12cm,angle=0]{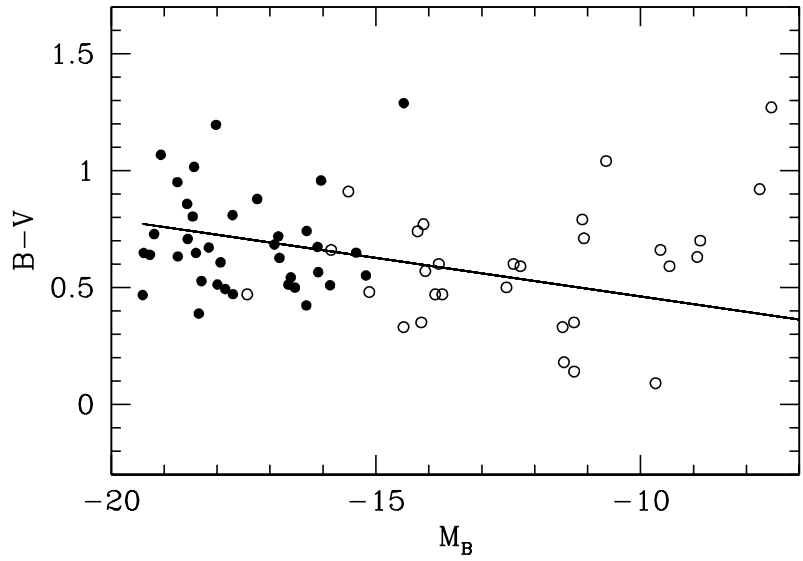}
\protect\caption[ ]{Color--magnitude diagram of the satellite galaxies
presented in this article ($filled\; circles$) and the dwarf galaxies of the
Local Group ($open \; circles$). The line is a least-squares fit to the
external early type satellite galaxies.} 
\label{compar_gl}
\end{center}
\end{figure}

We have compared the photometric properties of the satellites of our sample 
with those found for galaxies in the Local Group. The values for the
photometric integrated magnitudes have been taken from the compilation by
Mateo (1998) and correspond to all the members of the Local Group known by
that time, apart of the Milky Way, M31, and the two Magellanic clouds which
have been excluded from the analysis. This is illustrated in
Figure~\ref{compar_gl} in which we present the color--magnitude diagram for
both sets of objects. The typical uncertainties for our measurements in $B$ and $V$ are
$\sim$0.1 mag (when the colors are computed part of the
systematic errors cancel out). In this figure we have also
included  a few objects for which we do not have a morphological
classification and so excluded them from Figure~\ref{cmr}. For the Local Group
galaxies the uncertainties in the colors are $\sim$0.05 while, according to
the compilation by Mateo (1998), the uncertainty is, in general, large  for the
integrated magnitudes. Although we have not represented these errors, they are
not essential for making a qualitative analysis of the figure. We see how the
external satellites correspond to the brightest part of the Local Group
luminosity  function, and also to some brighter satellites with luminosities
similar to that of M33. This is expected because of the way in which the
original sample by Zaritsky et al.\ was built. Galaxies within the common
range of magnitudes have similar $B-V$ color and dispersions. The
least-squares fit of $B-V$ vs. $M_B$ obtained for the early-type external satellites  seems also
to fit notably well the galaxies of the Local Group. This extends the
validity of this color--magnitude relation through a range of 12 magnitudes
and demonstrates the universal nature of the color--magnitude relation for
satellite galaxies.

\section{Internal Color Gradients of Early-Type Satellites}

The existence of color gradients in early-type galaxies is well known.
 In general, elliptical galaxies tend to be redder in their inner regions.
As for the color--magnitude relation, the effect could be explained  by either a
gradient in stellar  age or metallicity, or by a combination of both. The
existence of dust more concentrated in the central parts of the galaxy could
also contribute totally or partially to the observed color gradients. The first
studies for giant ellipticals using CCD measurement were conducted by Franx \&
Illingworth (1990), Peletier et al.\ (1990), and de Jong (1996). Using the
{\itshape Hubble Space Telescope}, this study has been extended to galaxies  at higher
redshift  (Moth \& Elston 2002), who  show the existence
of negative gradients at intermediate ($0.5\le z\le 1.2$) redshifts, while the
gradients become very positive at higher ($2.0\le z\le 3.5$) redshifts. 

We have analyzed the gradients of the 16 galaxies classified as E or E/S0 in
the morphological analysis presented in Paper~I. Considering the angular size
of the objects and the limiting magnitude and spatial resolution of our data,
we decided to conduct a very simple analysis in the following way: i) we checked the value
of the seeing in both filters (in cases of significant different values, the
image in the filter with best seeing was convolved with a Gaussian of
appropriate width to match in the convolved image the seeing in the other
filter), ii) we selected the angular range between $2\times$ FWHM and the
radius which encloses the full galaxy and computed the integrated magnitudes in 
several concentric apertures within this range, and iii)  we did a linear
least-squares fit of $B-R$ vs. $\log r$. The slope of this relation is denoted
by $\Delta (B-R)/\Delta (\log r)$ and is presented in Table~3. In some cases we 
check the behavior of this gradient with that
obtained using other colors (usually $V-I$). 

A few galaxies were excluded from this analysis for the reasons detailed
below: NGC~895$a$, which has an off-center  nucleus and is also a rather compact object
with an angular size too small to compute the gradients;
NGC~2718$a$, which, apart from being an object in clear interaction with NGC~2718$b$,
has another object very close to it (we were able to separate it from the main
galaxy in the photometric analysis presented in this paper, but not in this
estimate of gradients); NGC~5962$a$, whose angular size is too small with
respect to the seeing; A910$a$ and NGC~2939$a$ because we  have observed
them in only one filter; and NGC~4541$b$,  which is in clear
interaction with NGC~4541$e$ and whose gradients shows a complex dependence on 
radius. We have also included  the galaxy NGC~3735$b$, classified
as early type in Paper~I, but which is not included in Table~1 because of the lack of
calibrated observations of this galaxy. After this selection, we were left with nine
objects, whose results are presented in Table~3. Seven of the nine galaxies have
negative slopes, while NGC~488$a$ has a value very close to 0 and, surprisingly,
NGC~4725$b$ shows an extremely positive gradients (i.e., the outer parts
are notably redder than  the inner ones. We have checked this trend
measuring the gradients in  other
colors and obtained the same tendency. Positive gradients such as this are
uncommon in low redshift galaxies and probably indicate strong star formation in
the central part of the galaxy. This galaxy is rather regular, and so far, we
have not found any clear evidence of distortions associated with any interactions
or mergers that could trigger this burst. The mean value of the
gradient (excluding NGC~4725$b$) is $\Delta (B-R)/\Delta (\log r)=-0.085\pm
0.027$.
\section{Summary}

\begin{enumerate}
\item We have presented integrated photometry of 49 satellite galaxies orbiting
external isolated spiral galaxies.

\item The $B-V$ vs.\ $M_V$ color magnitude relation is similar to the one 
found in the Fornax Cluster for galaxies with similar magnitudes. However 
the $V-I$ vs.\ $M_V$ relation is steeper in our sample as compared with
galaxies in Fornax.

\item The $B-V$ vs.\ $M_B$ color diagram extends to brighter magnitudes the
relations found for satellite galaxies in the Local Group.

\item We have measured the internal color gradients of nine early-type satellites.
Excluding a galaxy with a strong positive gradient, we found  
$\Delta (B-R)/\Delta (\log r)=-0.085\pm 0.027$. 
\end{enumerate}
\acknowledgments We would like to thank Francisco Prada for fruitful and
encouraging discussions. We also thank the anonymous referee for useful comments.

\clearpage

\onecolumn
\begin{longtable}{lllllll}
\caption{Integrated photometry of satellite galaxies. The columns are: 1)
name of the object; 2) recessional
velocity of the parent galaxy (from Zaritsky et al. 1997); 3-6)
$B$, $V$, $R$, and $I$ integrated magnitudes; telescope (1. IAC80, 2 NOT, 3. 1.23 m Calar Alto, 4.
2.2 m Calar Alto).} \\

Object & $V$ (km s$^{-1})$ & $B$ & $V$ & $R$ & $I$ & Telescope \\
\hline
\\
NGC~259b  	 &	   3808   &   15.04  & 14.37  & 13.83  & 13.30  &1\\
NGC~488a  	 &	   2268   &   15.39  & 14.48  & 13.93  & 13.34  &1\\
NGC~488b 	 &	   2268   &   16.18  & 15.63  & 15.29  & 14.99  &1\\
NGC~488c  	 &	   2268   &   16.60  & 15.61  & 15.22  & 14.62  &1\\
I1723a 	         &	   5531   &   19.45  & 18.85  & 17.99  & 18.10  &1\\
I1723b 	         &	   5531   &   16.94  & 16.08  & 15.53  & 14.91  &1\\
NGC~749a    	 &	   4406   &   -----  & -----  & 15.59  & -----  &1\\
NGC~749b    	 &	   4406   &   -----  & -----  & 16.93  & -----  &1\\
NGC~772a  	 &	   2468   &   14.42  & 13.49  & 12.86  & 12.19  &1\\
NGC~772b 	 &	   2468   &   -----  & -----  & 15.55  & 15.08  &1\\
NGC~772c 	 &	   2468   &   16.22  & 15.43  & 14.88  & 14.42  &1\\
NGC~895a   	 &	   2290   &   -----  & 16.72  & 16.49  & 16.13  &1\\
NGC~1517a 	 &	   3483   &   16.54  & 15.77  & 15.11  & 14.74  &1\\
NGC~1620a 	 &	   3513   &   -----  & 15.17  & 14.27  & -----  &1\\
NGC~1620b 	 &	   3513   &   14.91  & 13.91  & 13.26  & 12.74  &1\\
NGC~1961a         &	   3934   &   14.98  & 14.14  & 13.52  & 12.73  &4\\
NGC~1961b         &	   3934   &   15.82  & 14.88  & 14.25  & 13.38  &4\\
NGC~1961c         &	   3934   &   14.94  & 14.18  & 13.66  & 12.80  &4\\
NGC~1961d         &	   3934   &   14.81  & 14.22  & 13.51  & 12.95  &1\\
NGC~1961e  	 &	   3934   &   15.64  & -----  & 12.71  & 12.17  &1\\
NGC~2718a         &	   3842   &   17.32  & 16.71  & 16.44  & 16.14  &3\\
NGC~2718b         &	   3842   &   16.23  & 15.69  & 15.26  & 14.88  &3\\
NGC~2775a         &	   1357   &   15.48  & 14.87  & 14.59  & 14.28  &2\\
NGC~4162a         &	   2561   &   16.81  & 16.10  & 15.58  & 15.12  &2\\
NGC~4541a         &	   6898   &   16.45  & 15.72  & 15.11  & 14.43  &3\\
NGC~4541b         &	   6898   &   16.71  & 16.16  & 15.36  & 14.84  &3\\
NGC~4541d         &	   6898   &   17.07  & 16.44  & 16.03  & 15.65  &3\\
NGC~4541e         &	   6898   &   15.62  & 14.95  & 14.37  & 13.87  &3\\
A1242a           &	   6301   &   17.11  & 16.56  & 16.37  & 16.20  &2\\
NGC~4725a         &	   1207   &   13.01  & 12.33  & 11.88  & 11.27  &3\\
NGC~4725b         &	   1207   &   15.80  & 15.14  & 14.71  & 14.26  &2\\
NGC~5248a 	 &	   1154   &   15.25  & 14.72  & 14.57  & 14.34  &1\\
NGC~5324b          &	   3044   &   16.60  & 16.06  & 15.68  & 15.37  &2\\
A1416a   	 &	   6809   &   16.09  & 14.96  & 14.35  & 13.72  &2\\
NGC~5899a 	 &	   2563   &   14.43  & 13.39  & 12.67  & 11.95  &1\\
NGC~5921a 	 &	   1478   &   15.66  & -----  & 14.38  & -----  &1\\
NGC~5965a1        &	   3413   &   16.29  & 15.86  & 15.54  & 15.20  &2\\
NGC~5965a2        & 	   3413   &   17.19  & 16.50  & 16.60  & 16.29  &2\\
NGC~5962a 	 &	   1955   &   -----  & 17.50  & 16.92  & 16.08  &1\\
NGC~5962d 	 &	   1955   &   15.55  & 14.78  & 14.33  & 13.72  &1\\
NGC~6181a         &	   2371   &   14.46  & 13.75  & 13.23  & 12.69  &2\\
NGC~6384a 	 &	   1655   &   14.24  & 12.94  & 12.61  & 12.00  &1\\
NGC~7137a 	 &	   1691   &   -----  & -----  & 16.27  & 15.52  &1\\
NGC~7177a 	 &	   1150   &   14.56  & 13.85  & 13.22  & 12.55  &1\\
NGC~7184a 	 &	   2632   &   18.48  & 17.16  & 16.71  & 15.98  &1\\
NGC~7290a 	 &	   2900   &   17.00  & 16.51  & 16.05  & 16.17  &1\\
NGC~7290b 	 &	   2900   &   16.97  & 16.17  & 15.83  & 15.25  &1\\
NGC~7678a 	 &	   3486   &   15.30  & 14.86  & 14.35  & 13.94  &1\\
NGC~7755a 	 &	   2961   &   17.83  & -----  & -----  & 17.15  &1\\ 
\end{longtable}	

\newpage
\begin{table}
\caption{Color-magnitude relations for satellite
galaxies.}
\begin{center}
\begin{tabular}{cccccc}
Color & Type &$N$ & $a$ & $b$ & $\sigma$ \\
\hline
\\
$B-V$   & Early &   11 & $-0.048\pm 0.019$ &  $-0.158$& 0.140 \\
        & Late  &   22 & $-0.049\pm 0.012$ &  $-0.241$& 0.185 \\
$V-I$   & Early &   12 & $-0.138\pm 0.017$ &  $-$1.539 & 0.173 \\
        & Late  &   22 & $-0.193\pm 0.013$ &  $-$2.650 & 0.249 \\
\end{tabular}
\end{center}
\end{table}

\newpage

\begin{table}
\caption{Color gradients for early-type satellite galaxies.}
\begin{center}
\begin{tabular}{lcc}
Galaxy & Range  & $\Delta (B-R)/\Delta (\log r)$ \\
     & (arcsec) & (mag arcsec$^{-2}$) \\
 \hline
 \\
NGC~259b & $2-8$ & $-0.031$ \\
NGC~488a & $2-6$ & +0.009 \\
I1723b   & $2-6$ & $-$0.056 \\
NGC~772a & $2-7$ & $-$0.076 \\
NGC~1620b& $2-7$ & $-$0.121 \\
NGC~2718b& $2-5$ & $-$0.216 \\
NGC~3735b& $2-7$ & $-$0.169 \\
NGC~4725b& $2-7$ & +0.621 \\
NGC~7290b& $2-4$ & $-$0.023 \\
\end{tabular}
\end{center}
\end{table}     

\clearpage
\newpage

\newpage

\newpage

\newpage
 
\end{document}